# Conditional entropy and weak fluctuation correlation in nonequilibrium complex systems

Yuichi Itto[1,2]

[1] Science Division, Center for General Education, Aichi Institute of Technology,
 Aichi 470-0392, Japan

[2] ICP, Universität Stuttgart, 70569 Stuttgart, Germany

**Abstract.** The weak correlation between spatiotemporal fluctuations in nonequilibrium complex systems is shown to govern the fluctuation distribution, maximizing the conditional entropy associated with such fluctuations. The result is illustrated in diffusion phenomena observed in living cells. A generic feature of the weak correlation is briefly mentioned.

# 1. Introduction

There exists a wide class of complex systems globally in nonequilibrium-stationary-state-like situation in the sense that each of its small spatial regions is in a local equilibrium state characterized by dynamics of two different fluctuating quantities, e.g., the local temperature, on a large time scale: the time scale of their dynamics is much larger than that of a typical local dynamics, e.g., the one of a random walker. Denoting such quantities by $x$ and $y$, which are random variables, the joint probability distribution is given here by $g(x,y) = g(x|y) f(y)$, where $g(x|y)$ is the conditional distribution describing the probability of $x$, given a value of $y$, and $f(y)$ is the marginal distribution describing the probability of $y$. At a certain given value of $y$, say $y_0$, the conditional entropy associated with $g(x|y_0)$ in the form of the Shannon entropy [1] is as follows:

$$S[g] = -\int dx\, g(x|y_0) \ln g(x|y_0). \qquad (1)$$

The maximum entropy principle [2] has been used as the crux for treating nonequilibrium complex systems with a hierarchical structure of different dynamics on different time scales, as in recent works, e.g., in Refs. [3-10], where the joint entropy associated with the fast and slow dynamics or the marginal entropy concerning the slow dynamics is maximized. For the joint entropy, the marginal distribution is of central importance while the conditional distribution is well understood by local equilibria, there. In contrast to this, there are often situations, where the latter is the distribution of relevance, as will be seen later. It may be worth pointing out that the maximum entropy principle has played a crucial role for studying dynamical systems similar to those in the field of social sciences [11] (see, e.g., Ref. [12]).

Maximization of such entropies seems to be reasonable since systems are on largely separated time scales. A point in our present context is as follows. Recalling that the system under consideration is in nonequilibrium-stationary-state-like situation, each of the local regions mentioned above is in a quasi-equilibrium state with fluctuating local temperature to be described by a canonical ensemble. The stationarity is considered to hold due to the time scale separation: the conditional entropy in Eq. (1) is maximized in this sense.

For Eq. (1), a further consideration, which seems to be necessary, is as follows. The local regions of the original system are regarded as imaginary blocks that are independent each other with respect to $x$, allowing one to construct a lot of collections formed by the blocks. The statistical property of the conditional fluctuations in a given collection is then equivalent to that in the original system but the local property is not. So, a measure about uncertainty of the conditional fluctuations is introduced in a manner similar to the one for deriving the Shannon entropy [2], by which the measure is expected to become the entropy in Eq. (1) (see also Refs. [7-9]).

In this article, we report a conditional entropic approach developed in Ref. [13] for studying nonequilibrium complex systems with a weak correlation between slowly fluctuating quantities. The weak correlation is found to govern the conditional fluctuation distribution that maximizes the entropy in Eq. (1). The result is illustrated in diffusion phenomena in living cells. In addition, we briefly mention a generic structure of the weak correlation.

## 2. Weak correlation and maximum entropy principle

Following Refs. [14,15], the weak correlation is given by the fluctuations of $x$ and $y$ that are not fully statistically independent each other. Let us formally write $g(x|y) = e^{h(x|y)}$ with a suitable function, $h(x|y)$, for which the following expansion holds in the whole range of $y$: $h(x|y) \cong h_0(x) + h_1(x)(y - y_0)$ with $h_0(x) \equiv h(x|y_0)$ and $h_1(x) \equiv h'(x|y_0)$, where $y_0$ is taken to be the average value of $y$ and the prime denotes differentiation with respect to $y$. So, $h_1(x)$ should be small so that the function is approximately constant in terms of $y$. In other words, the degree of correlation between $x$ and $y$ is small, i.e., weak. This weakness seems to reflect the nonequilibrium-stationary-state-like situation, which is far from the strongly nonequilibrium regime. There, the heat flux created by the temperature gradient between the local regions is considered to exist, in general, which may cause weakly correlated regions.

Thus, the conditional distribution is given by

$$g(x|y) \sim g(x|y_0) \exp[(y - y_0) h_1(x)], \tag{2}$$

which offers the marginal distribution $g(x) = \int dy\, g(x, y)$ calculated to be

$$g(x) \sim g(x|y_0) \exp[-y_0 h_1(x)] \int dy\, f(y) \exp[h_1(x) y]. \tag{3}$$

It has been shown in Refs. [14,15] that the existence of the weak correlation is essential for describing the marginal distribution. As simple examples, the weak correlation has been demonstrated in Ref. [13] for the bivariate exponential distribution [16] and the bivariate Gaussian distribution [17].

In what follows, we see that $g(x|y_0)$ is determined in terms of the weak correlation $h_1(x)$ by the maximum entropy principle with the conditional entropy in Eq. (1): the $x$ dependence of $h_0(x)$ appears only from $h_1(x)$ *itself*, except possible additional quantities irrelevant to the weak correlation.

To see this, let us assume that $h_1(x)$ is known. Then, under the expansion of the exponential factor in Eq. (2) as well as the normalization condition on $g(x|y_0)$, the normalizability of $g(x|y)$ tells us that the following condition has to be naturally imposed:

$$\int dx\, g(x|y_0) h_1(x) = 0, \qquad (4)$$

which is reasonable since the first-order term with respect to $h_1(x)$ is dominant in the expansion, whereas the higher-order terms are not, recalling the smallness of $h_1(x)$. Clearly, this condition means that the average of $h_1(x)$ should vanish (see Section 5).

Therefore, together with a possible constraint on the average of a certain quantity,

$Q(x)$, the maximum entropy principle reads

$$\delta_g \left\{ S[g] - \lambda \left( \int dx\, g(x|y_0) - 1 \right) + \nu \left( \int dx\, g(x|y_0) h_1(x) - 0 \right) \right. $$
$$\left. + \kappa \left( \int dx\, g(x|y_0) Q(x) - \overline{Q} \right) \right\} = 0, \qquad (5)$$

where $\lambda$, $\nu$, and $\kappa$ are the set of the Lagrange multipliers associated with the constraints on the normalization condition, the weak correlation, and the average value, respectively, and $\delta_g$ denotes the variation in terms of $g(x|y_0)$. The stationary solution of Eq. (5) is as follows [13]:

$$g^*(x|y_0) \propto \exp[\nu h_1(x) + \kappa Q(x)], \qquad (6)$$

showing that the conditional distribution *at* $y = y_0$ is, in fact, realized by the weak correlation.

## 3. DNA-binding proteins in bacteria

As shown in Ref. [13], we illustrate our conditional entropic approach in diffusion phenomena observed in a recent experiment in Ref. [18], which has revealed that histonelike nucleoid-structuring proteins known as DNA-binding proteins exhibit heterogeneous diffusion with fluctuations in *Escherichia coli* bacteria at the level of their individual trajectories.

The diffusion property is characterized by the mean square displacement for large

elapsed time, $t$:

$$\overline{\Delta x^2} \sim D_\alpha t^\alpha, \qquad (7)$$

where $D_\alpha$ is the diffusion coefficient showing the asymptotic power-law distribution given by

$$\varphi(D_\alpha) \sim D_\alpha^{-\gamma-1} \qquad (8)$$

with $\gamma \cong 0.97$, whereas $\alpha$ is the diffusion exponent obeying a non-trivial distribution in the wide range $0 \leq \alpha \leq 2$, see Figs. 2d and 2c, respectively, in Ref. [18]. Note that the distribution in Eq. (8) is obtained for dimensionless numerical values of $D_\alpha$. Normal diffusion is the case of $\alpha = 1$, while the case of $\alpha \neq 1$ is referred to as anomalous diffusion [19] (see Refs. [20-22] for reviews). At small elapsed time, only normal diffusion has been observed, the diffusion coefficient, $D$, of which asymptotically follows a power law.

A crucial point is the fact [18] that the average diffusion exponent increases *only slightly* with respect to the cell age (or, equivalently, cell length) in marked contrast to a significant increase of the average diffusion coefficient, both of which are obtained from the analysis of the mean square displacement in an ensemble average, i.e., an average of square displacement over all of the individual trajectories. In Ref. [14], it has therefore been suggested, based on the assumption of the Einstein relation [23], $D \propto 1/\beta$ with the inverse temperature, $\beta$, that the correlation between the diffusion-exponent fluctuations and the temperature fluctuations is weak.

Accordingly, in our present context, $x$ and $y$ are taken to be $\alpha$ and $\beta$, respectively.

To examine the conditional distribution $g(\alpha|\beta)$, the following inverse gamma distribution has been considered as a concrete form for Eq. (8) [14]:

$$\varphi(D_\alpha) \propto \tilde{A}^\gamma D_\alpha^{-\gamma-1} \exp\left(-\frac{\gamma \tilde{A}}{D_\alpha}\right) \qquad (9)$$

with a dimensionless positive constant, $\tilde{A}$, in an interval, which decays as the power law in Eq. (8) and fits the experimental data well, see Fig. 1 in Ref. [14]. $D_\alpha$ has been assumed to depend also on $\beta$, i.e., $D_\alpha = D_{\alpha,\beta}$ in a way similar to the Einstein relation [24]:

$$D_{\alpha,\beta} \sim \frac{c}{s^\alpha \beta}, \qquad (10)$$

where $s$ denotes a typical time characterizing the displacement of the protein and $c$ is a positive constant. It has been then proposed that given a value of $\beta$, $D_\alpha$ follows an inverse gamma distribution in Eq. (9) with $\tilde{A} = \tilde{A}(\beta)$. So, from $g(\alpha|\beta) = |\partial D_{\alpha,\beta}/\partial \alpha| \varphi(D_\alpha)$, the conditional distribution in this case is found to be

$$g(\alpha|\beta) \propto s^{\gamma\alpha} \exp\left[-\frac{\gamma a(\beta)}{c} s^\alpha\right], \qquad (11)$$

where $\tilde{A}(\beta) = a(\beta)/\beta$ with a positive quantity, $a(\beta)$, has been used. To realize the weakness of correlation, $a(\beta)$ is expanded around at $\beta = \beta_0$, which is the average value of $\beta$, in such a way that $a(\beta) \cong a_0 + a_1(\beta - \beta_0)$ with $a_0 \equiv a(\beta_0)$ and $a_1 \equiv a'(\beta_0)$: $a_1$ is small and is negative in accordance with the cell-age dependence [14].

Thus, the weak correlation is obtained as

$$h_1(\alpha) = \frac{\gamma a_1}{c}(\langle s^\alpha \rangle_\alpha - s^\alpha) \tag{12}$$

with $\langle \bullet \rangle_\alpha$ being the average with respect to $g(\alpha | \beta_0)$. Using $\chi^2$ distribution for $f(\beta)$ [25], the marginal distribution in Eq. (3) with Eqs. (11) and (12) has been obtained in good agreement with the experimental data, see Fig. 2 in Ref. [14].

As mentioned above, the mean square displacement in Eq. (7) has been analyzed at the level of individual trajectories. This fact seems to suggest that the imaginary blocks mentioned in the Introduction are independent each other in terms of $\alpha$. This, in turn, allows us to employ the present entropic approach based on Eq. (1). Taking $Q(\alpha) = \alpha$, therefore, we have the following conditional distribution:

$$g^*(\alpha | \beta_0) \propto \exp[\nu h_1(\alpha) + \kappa \alpha], \tag{13}$$

which becomes equivalent to the distribution in Eq. (11) after the following choices are

made:

$$v = \frac{a_0}{a_1}, \qquad \kappa = \gamma \ln s, \qquad (14)$$

provided that all quantities appearing here are dimensionless.

## 4. Membraneless organelles in embryos and beads in cell extracts

We further develop a possible illustration [13] based on the gross behavior of fluctuations observed experimentally in diffusion phenomena in two different systems, which share common natures both in the diffusion-coefficient fluctuations and the diffusion-exponent fluctuations. One is of the p-granules, which are membraneless organelles, in embryos of *C. elegans* in Ref. [26], and the other is the beads in cell extracts obtained from the eggs of *Xenopus laevis* in Ref. [27].

Like in Eq. (7), not only $\alpha$ but also $D_\alpha$ in the context of these systems fluctuates in a wide range. The distribution of the former is of a unimodal form, see Fig. 2a in Ref. [26] and Fig. 2a in Ref. [27], whereas the distribution takes almost a log-normal form for the latter, see Fig. 2b in Ref. [26] and Fig. 2b in Ref. [27], for which the values of $D_\alpha \times 1\mathrm{s}^\alpha$ having the dimension of spatial area have been employed, given a value of $\alpha$. In what follows, we use the same notation $D_\alpha$ as the values of $D_\alpha \times 1\mathrm{s}^\alpha$ for the sake of simplicity.

Therefore, again, we suppose under the assumption of Eq. (10) that $x$ and $y$ correspond to $\alpha$ and $\beta$, respectively.

The distribution of $D_\alpha$ is of a log-normal form in an interval:

$$\varphi(D_\alpha) \propto \frac{1}{D_\alpha} \exp\left\{-\frac{\left[\ln\left(D_\alpha / \tilde{A}\right)\right]^2}{2m^2}\right\}, \tag{15}$$

where $\tilde{A}$ in the present case and $m$ are positive constants. Let us assume that the distribution of $D_\alpha$ with a given value of $\beta$ also takes a log-normal type in Eq. (15) with $\tilde{A} = \tilde{A}(\beta)$. In the same procedure in Section 3, therefore, the Gaussian form of the conditional distribution is given by

$$g(\alpha | \beta) \propto \exp\left\{-\frac{\left[\alpha \ln s - \ln\left(c / a(\beta)\right)\right]^2}{2m^2}\right\} \tag{16}$$

in the range $0 < \alpha \leq \alpha_{max}$, where $s$ has a certain value less than unity and $\alpha_{max}$ is the maximum value of $\alpha$ [26,27], provide that $\tilde{A}(\beta) = a(\beta) / \beta$ with the same expansion for $a(\beta)$. It is noted that the condition, $c / a_0 < 1$, is fulfilled in order for $g(\alpha | \beta_0)$ to have a peak.

Thus, the weak correlation in this case is as follows:

$$\begin{aligned} h_1(\alpha) &= \frac{a_1}{a_0}\left[\frac{\ln(c / a_0) - \alpha \ln s}{m^2} + \frac{g(0 | \beta_0) - g(\alpha_{max} | \beta_0)}{\ln s}\right] \\ &= \frac{a_1}{m^2 a_0}\left[\ln\left(\frac{c}{a_0}\right) - \alpha \ln s\right], \end{aligned} \tag{17}$$

where $g(0 | \beta_0) = 0$ and $g(\alpha_{max} | \beta_0) = 0$ have been used at the second equality in consistent with the experimental results, see Fig. 2c in Ref. [26] and Fig. 2c in Ref. [27]. It has been shown in Ref. [13] that based on a log-normal distribution for $f(\beta)$, the same trend of the unimodality of the diffusion-exponent fluctuation distribution [26,27] is recognized by the marginal distribution in Eq. (3) with Eqs. (16) and (17) under the condition, $(m^2/\sigma^2)(a_0/a_1)^2 > 1$, where $\sigma^2$ is the variance of $\beta$ with respect to $f(\beta)$.

The experimental facts [26,27] that both of the trajectories of the p-granules and the beads are individual again justify the use of the entropy in Eq. (1), like in Section 3. Accordingly, taking $Q(\alpha) = \alpha^2$, which imposes a constraint on the size of the fluctuations of $\alpha^2$, (see also Ref. [8]), we have the weak correlation in this case as follows:

$$g^*(\alpha | \beta_0) \propto \exp[\nu h_1(\alpha) + \kappa \alpha^2], \qquad (18)$$

for which it turns out that this becomes identical to the conditional distribution in Eq. (16) under the following choices:

$$\nu = \frac{a_0}{a_1} \ln\left(\frac{a_0}{c}\right), \qquad \kappa = -\frac{(\ln s)^2}{2m^2}. \qquad (19)$$

## 5. Generic feature of weak correlation

In this section, we briefly mention a generic feature of the weak correlation. From Eq. (4) with $g = g^*$, the weak correlation is written, without loss of generality, by

$$h_1(x) = z(x) - \langle z(x) \rangle_x^*, \tag{20}$$

where $z(x)$ is a quantity depending on $x$, given $y_0$, and $\langle \bullet \rangle_x^*$ stands for the average with respect to $g^*(x | y_0)$, (and it is understood that the Lagrange multipliers will be appropriately chosen). With this, it is obvious that the condition in Eq. (4) is in fact satisfied. Regarding the part of the weak correlation, therefore, the conditional distribution in Eq. (6) is seen to be realized by $z(x)$ [since the second term in Eq. (20) can be included in the normalization constant].

In the illustrative examples in Sections 3 and 4, $z(\alpha)$ is naturally taken as follows: $z(\alpha) = -(\gamma a_1 / c) s^\alpha$ in the case of DNA-binding proteins, whereas $z(\alpha) = -[(a_1 \ln s)/(m^2 a_0)]\alpha$ in the case of the p-granules as well as the beads. Thus, these quantities are expected to be essential for studying the conditional fluctuations observed in diffusion phenomena discussed here.

## 6. Concluding remarks

We have reported a conditional entropic approach to nonequilibrium complex systems with a weak correlation between slowly fluctuating quantities on a large time scale. The approach has shown that the fluctuation distribution, which maximizes the conditional entropy concerning the fluctuations, is organized by the weak correlation. The result has been demonstrated in diffusion phenomena in living cells. A brief discussion has also been made about a generic structure of the weak correlation.

We make some comments on the following. Heterogeneous diffusion of DNA in the cell nuclei in a human osteosarcoma cell line have been reported in a recent experimental work in Ref. [28] (see also, e.g., Refs. [29,30] for relevant issues). In a certain population of the diffusion coefficient as well as the diffusion exponent, the former is log-normally distributed, whereas the latter is Gaussian-distributed, see Figs. 2d and 2e in Ref. [28], respectively. Then, the dimensionless values of the diffusion coefficient have also been evaluated, the distribution of which has been found to be a log-normal form, see Fig. S4g in Ref. [28]. Therefore, from the discussions similar to those in Section 4 in terms of the weak correlation as well as the conditional distribution, it is expected that the marginal distribution of the diffusion-exponent fluctuations is close to the Gaussian form, if the distribution of the temperature fluctuations is sharply peaked around its average.

The weak correlation between the diffusion exponent and temperature may be found in a recent experiment, e.g., in Ref. [31], where a mild dependence of the diffusion exponent on temperature has been observed in diffusion of telomeres inside cells.

The role of spatiotemporal fluctuations in heterogeneous diffusion is a basic issue in connection with time evolution within the framework of so-called superstatistics [32], e.g., in Refs. [33-41]. A similar issue has also been mentioned in Ref. [42] for an epidemic model, which may suggest its relevance to the present approach.

**Acknowledgement**

This work is supported by a Grant-in-Aid for Scientific Research from the Japan Society for the Promotion of Science (No. 21K03394).